\documentclass[aps,prl,twocolumn,showpacs,groupedaddress]{revtex4}
\usepackage{graphicx}  
\usepackage{dcolumn}   
\usepackage{bm}        
\usepackage{amssymb}   
\usepackage{subfigure}
\usepackage{amsmath,amsfonts}
\usepackage{setspace}
\usepackage[mathscr]{euscript}

\begin{document}
\title{Topological Transition to a Critical Phase in a Two-dimensional 
3-Vector Model\\
 with non-Abelian Fundamental Group: A Simulational Study
}

\author{B. Kamala Latha$^{1}$}
\author{V.S.S. Sastry$^{2}$}
\affiliation {$^{1}$School of Physics,University of Hyderabad, Hyderabad 500046, India}

\affiliation{$^{2}$Centre for Modelling, Simulation and Design, University of Hyderabad, Hyderabad 500046, India}

\date{\today}%

\begin{abstract}
Two-dimensional 3-vector (\textit{d}=2, \textit{n}=3) lattice model with 
inversion site symmetry and fundamental group of its order-parameter 
space $\Pi_1 (\mathcal{R})= Z_{2}$, did not exhibit the expected 
topological transition despite  stable defects associated with 
its uniaxial orientational order. This model is investigated specifically 
requiring the medium to host distinct classes of defects associated with 
the three ordering directions, facilitating their simultaneous 
interactions. The necessary non-Abelian isotropy subgroup of 
$\mathcal{R}$ is realized by assigning $D_{2}$ site symmetry, resulting 
in $\Pi_1 (\mathcal{R})= \mathbb{Q }$ (the group of quaternions). With 
liquid crystals serving as prototype model, a general biquadratic 
Hamiltonian is chosen to incorporate equally attractive interactions 
among the three local directors resulting in an orientational order with 
the desired topology. A Monte Carlo investigation based on the density 
of states shows that this model exhibits a transition, simultaneously 
mediated by the three distinct defects with topological charge $1/2$
(disclinations), to a low-temperature critical state characterized by 
a line of critical points with quasi-long range order of its directors, 
their power-law exponents vanishing as temperature tends to zero. It 
is argued that with \textit{n}=3, simultaneous participation of all 
spin degrees through their homotopically inequivalent defects is 
necessary to mediate a transition in the two-dimensional system to 
a topologically ordered state.

\end{abstract}

\pacs{64.70.M-,64.70.mf}
\maketitle

In 2-dimensional lattice systems with 3-dimensional ‘spin’ degrees of 
freedom, (\textit{d}=2,\textit{n}=3)  models, symmetry of the Hamiltonian
 impacts its order parameter space ($\mathcal{R}$)  topology \cite{Mermin},
 requiring non-trivial first fundamental group $\Pi_1 (\mathcal{R}$) to 
 sustain stable topological point defects. With a choice of minimal local 
 site symmetry $Z_{2}$, $\mathcal{R}$ is isomorphic to $ RP^{2}$ 
 (3-d real projective space, $\Pi_{1}(\mathcal{R}) = Z_{2}$)  
 resulting in an apolar uniaxial order (in the direction of symmetry), 
 forming stable point-defects (disclinations). 
 Monte Carlo (MC) studies on such models could not establish the 
  presence of a Berenzskii-Kosterlitz-Thouless-type (BKT) transition
 mediated by topological defects \cite{Berz2, Kosterlitz}, with examples
  from liquid crystals (LC)\cite{Kunz, BP,Mondal, Dutta,Shabnam} and 
 magnetic systems \cite{Kawamura}. In a related MC study based on 
 density of states \cite{BKLPRL}, it was observed that the initial 
 progression of the system towards such a transition was interrupted 
 by a crossover arising from competing length scales in the system. 
We investigate this system by requiring that its order-parameter 
topology allows for a (discrete) non-Abelian fundamental group 
resulting in distinct classes of topological defects associated 
with all the spin degrees.   
  
We assign $D_{2}$ symmetry to the lattice sites (instead of $D_{\infty h}$
 of the earlier model), and augment the Lebwohl-Lasher (LL) Hamiltonian
 \cite{LL} (representing an attractive biquadratic interaction among, 
 say, molecular z-axes), with similar attractive biquadratic interactions
  among the molecular x-axes and y-axes.  The corresponding $\mathcal{R}$ 
 is represented by the space of cosets $SU(2)/\mathbb{Q}$,
 where $SU(2)$ is the special unitary group  of $2 \times 2$ 
 matrices and $\mathbb{Q}$ is the discrete non-Abelian group of 
 quaternions. In this case, 
 $\Pi_{1}(\mathcal{R})= \mathbb{Q}$, represented by ($\bm{\pm 1},
 \bm{\pm i\sigma_{x}}, \bm{\pm i\sigma_{y}},\bm{\pm i\sigma_{z}}$); 
 $(\bm{\sigma_{i} })$ is the  set of Pauli matrices. The higher order 
 groups are not relevant to 2d models. The
 medium hosts four types of stable topological defect 
 structures: three distinct types of disclinations (charge $1/2$),
 corresponding to the order directors associated with the three 
 molecular axes and homotopically equivalent topological
 defects of unit charge formed by each of the axes \cite{Mermin}. 
 
 We define two orthogonal (uniaxial and biaxial) molecular tensors,
  $\bm{q}$ and  $\bm{b}$ respectively, as 
 $\bm{q} := \bm{m} \otimes \bm{m} - \frac{\bm{I}}{3}$ and
$\bm{b} := \bm{e} \otimes \bm{e} - \bm{e}_{\perp} \otimes \bm{e}_{\perp}$
where $(\bm{e},\bm{e_{\perp}},\bm{m})$ is an orthonormal set of vectors 
representing the molecular axes (in conventional notation \cite{Sonnet}). 
 The general biquadratic attractive interaction between 
 two lattice sites is given by 
$ H=-U[\xi \, \bm{q} \cdot \bm{q}^{\, \prime}
+ \gamma(\bm{q} \cdot \bm{b}^{\, \prime} + \bm{q^}
{\, \prime} \cdot \bm{b}) + \lambda \, \bm{b} \cdot \bm{b}^{\, \prime}]$.
This Hamiltonian, setting $\xi = 1$, was extensively examined in three
dimensional systems in the parameter space of ($\gamma, \lambda$), to
elucidate its phase diagram \cite{Sonnet, Bisi, BKL15,
BKL18}. We set $\gamma$ = 0 hereafter to avoid cross-coupling 
interactions, without loss of generality. \textit{H} 
can be expressed in terms of inner products of the molecular axes 
$(\bm{e},\bm{e_{\perp}},\bm{m})$, indexing them as (1,2,3) for convenience. 
The  pair-wise interaction between two lattice sites ($\alpha,\beta$)
 then simplifies to 
$H_{\alpha\beta}= - U \lbrace \xi \ G_{33} +
\lambda \ [2(G_{11}+G_{22})-G_{33}]\rbrace.$
Here $G_{ij}$ = $P_{2}$($f_{ij}$), $P_{2}(.)$ denoting the second 
Legendre polynomial and
 $f_{ij}$= ($\bm{u}_{i} \cdot\bm{v}_{j}$) where ($\bm{u}_{i}, i = 1,2,3)$
 and ($\bm{v}_{j}, j$ = 1,2,3) are the two triads of molecular axes 
 on the sites ($\alpha,\beta$) respectively \cite{Romano}. 
 Reduced temperature ($T$) is defined in units of $U$.
 With the choice of model parameters $\xi$=1  and $\lambda=\frac{1}{3}$
 the model exhibits cyclic permutation symmetry with respect to the
 indices of the local directors, imparting equally attractive interaction
 among the three axes. This choice leads to the strongest first-order 
 transition directly from isotropic (disordered) phase to a state with
  three ordering directions (biaxial phase) \cite{Bisi}.

 The MC simulations are carried out adopting  the Wang-Landau 
 algorithm, to calculate the density of states of the system, and hence 
 extract equilibrium averages of different physical properties of interest 
 (as described in \cite{WL,jayasri05, BKL15,BKL18})
 at $10^{3}$ temperatures over the range [0.1 to 1.5].
  We consider interactions of the nearest neighbour sites on 2d 
  square lattices with different sizes $L \times L$, ($L$ = 60, 80, 100, 
 120,150) and apply periodic boundary conditions. The computed properties 
 include the averages of energy per site ($E$), the specific heat $C_{v}$, 
 the uniaxial ($R_{00}^{2}$) and biaxial ($R_{22}^{2}$)  orientational 
 order parameters, as well as their susceptibilities $\chi_{00}^{2}$  
 and $\chi_{22}^{2}$ \cite{BKL15}. In addition, we also computed the 
 topological parameters of the dominant charge $1/2$ defects 
 of the three order directors. The fourth permissible charge 1 defect 
 is difficult to be detected due to energetic reasons \cite{Hind,Dutta}.
 
 The topological order parameter $\mu_{z}$, of the z-axis 
  director forming the charge $1/2$ defect,
  is calculated   by assigning a unit vector $\bm{s(r)}$ at 
  each site $\bm{r}$ on the square lattice  representing the local 
  \textit{z}-director orientation. For each bond  
  $(\bm{r}, \bm{r}^{'}$) the  shortest geodesic connecting
  the vectors $\bm {s(r)}$  and  $\bm {s(r^{'})}$ 
  on the unit (\textit{n}-1)-sphere (\textit{n} = 3) is chosen, thus
  obtaining a map for a closed loop  $\mathcal{L}$ on the lattice to 
  a loop on the manifold  $RP^{2}$ of  this director. The homotopy 
  class of this map is given by
$\mathcal{W(L)}= \prod_{(\bm{r},\bm{r^{'}}) \in \CMcal{L}}~ 
{sgn}(\bm{s(r)}, \bm{s(r^{'})})$, the product being 
sequentially ordered over $\mathcal{L}$. Topological order $\mu_{z}$
 is computed as the ensemble average of $\mathcal{W(L)}$  with periodic 
 boundary conditions in place, and a related parameter is calculated as
 $\delta_{z}= (1-\mu_{z})/2$ \cite{Kunz}. 
 We computed the density of  unbounded charge 
 $1/2$ defects $d_{z}$ of the director  connected 
 with the \textit{z}-axis, by dividing the 
 lattice into a composition of elementary triangular plaquettes.
 The above product applied to each plaquette yields a defect finding 
 algorithm: if the ordered product is -1, the plaquette 
 encloses a charge $1/2$ defect.  The average defect density 
 $d_{z}$ is calculated from the total count of such isolated 
 defects over the lattice and averaged over the ensemble \cite{Dutta}. 
 The topological parameters of the other two directors 
 ($\delta_{x}, \delta_{y} ; d_{x}, d_{y})$ 
 are similarly computed. Pair correlation functions of the spatial 
 variation of reorientational fluctuations, 
 $G(r_{ij})= < P_{2} \ (\cos \theta_{ij})>$, 
 are computed for the three directors (at \textit{L} = 150) at about 80
 temperatures. Statistical errors, estimated based on the  Jack-knife 
 algorithm \cite{BKL16}, in $E$, $R^{2}_{00}$,$R^{2}_{22}$,
  $ \delta_{(x,y,z)}$, and $ d_{(x,y,z)}$ are typically of the order 
  of 1 in $10^{4}$, while higher moments ($C_{v}$, $\chi^{2}_{00}$, 
  $\chi^{2}_{22}$) are relatively less accurate (about 8 in $10^{3}$). 
 \begin{figure}
\centering
\includegraphics[width=0.45\textwidth]{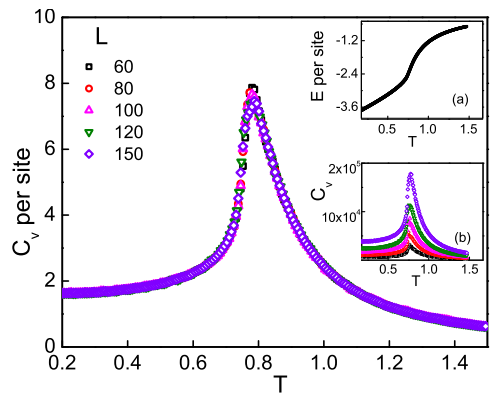}
\caption{(color online) Temperature variation of specific heat 
(per site) at lattice sizes  \textit{L} = 60, 
80, 100, 120, 150. Insets show the temperature variation of 
(a) size independent energy per site; (b) size dependence of $C_{v}$ 
for different \textit{L}. } 
\label{fig:1}
\end{figure} 

\begin{figure}
\centering
\includegraphics[width=0.45\textwidth]{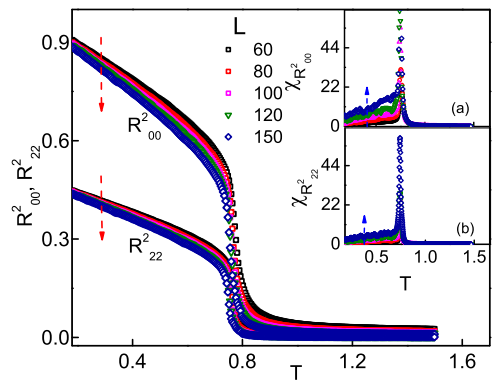}
\caption{(color online) Temperature variation of orientational order 
parameters at lattice sizes  \textit{L} = 60, 
80, 100, 120, 150. Insets show the temperature variation of 
(a) uniaxial susceptibility; (b) biaxial susceptibility for different
\textit{L}. (The arrows indicate increase in the size of the system). } 
\label{fig:2}
\end{figure}   
  
Fig.~\ref{fig:1} depicts the temperature variation of $C_{v}$ 
(per site) which is independent of the system size, unlike in 
a normal order-disorder transition. The two
insets show the size dependence of $C_{v}$ and the temperature variation
of the energy (per site) which is found size independent 
(within errors). Fig.~\ref{fig:2} shows the size dependence
 of $R^{2}_{00}$ and $R^{2}_{22}$, plotted 
as a function of temperature, and  both the order parameters 
decrease with increase in size. Also the temperatures of their onset 
are coincident at a given size, shifting to lower values 
with increase in size. The corresponding susceptibilities ($\chi_{00}^{2}$,  
 $\chi_{22}^{2}$), suitably scaled to the system sizes, are depicted 
 in the two insets. 
 The size dependence of peak locations of ($\chi_{00}^{2}$,  
 $\chi_{22}^{2}$) are commensurate with the corresponding dependence of 
 onsets of their orders. Their low temperature values below the 
 transition temperatures show a progressive divergence with size
  (insets of Fig.~\ref{fig:2}). These features of the orientational
  orders are typical signatures of a topologically ordered medium
  \cite{Mondal, Botet}.           
   \begin{figure}
\centering
\includegraphics[width=0.45\textwidth]{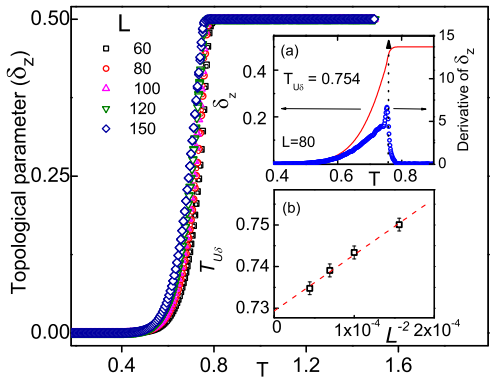}
\caption{(color online) Temperature variation of topological  
parameter $\delta_{z}$ at lattice sizes  \textit{L} = 60, 
80, 100, 120, 150. Inset (a) shows inflexion point of $\delta_{z}$ at 
\textit{L} = 80 (dashed vertical line) indicating the unbinding 
transition temperature $T_{U\delta}$; (b) Finite size scaling  plot 
of $T_{U\delta}(L)$.} 
\label{fig:3}
\end{figure}   

The permutation symmetry of the model implies identical variation 
of  the topological parameters ($\delta_{x},\delta_{y},\delta_{z}$) 
and the defect densities ($d_{x}, d_{y}, d_{z}$) 
with temperature at a given size, as the present observations confirm.
Fig.~\ref{fig:3} depicts the size-dependence of the temperature profiles 
of $(\delta_{x},\delta_{y}, \delta_{z})$
showing gradual shift to lower temperatures 
with increase in size, - much like  the orientational order
profiles and their susceptibility peak positions (see Fig.~\ref{fig:2}). 
The inflexion point of  the topological order parameter $\delta$
with respect to temperature corresponds to the unbinding transition 
temperature $T_{U\delta}(L)$ and the inset (a) of Fig.~\ref{fig:3} 
depicts its temperature derivative at L=80, with
 a peak at $T_{U\delta}$(L)=0.754 ($\pm$ 0.002). Inset (b) of 
 Fig.~\ref{fig:3} is a finite-size scaling plot of such transition 
 temperatures derived from topological parameter profiles at different 
 sizes, resulting in a reasonable fit and
 yielding an estimate of the unbinding temperature from this parameter 
 as $T_{U\delta}$=0.727 ($\pm$0.002) in the thermodynamic limit.  
\begin{figure}
\centering
\includegraphics[width=0.45\textwidth]{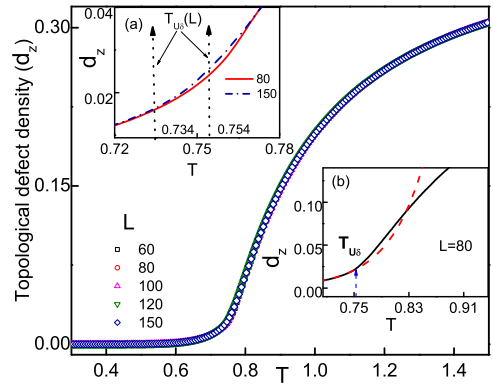}
\caption{(color online) Temperature variation of the topological 
defect density $(d_{x}, d_{y}, d_{z})$ at lattice sizes  
\textit{L} = 60, 80, 100, 120, 150. Inset (a) The temperature 
variations of $d_{z}$ show their size dependence near critical
region (shown at \textit{L} = 80, 150). The dotted vertical lines 
indicate the values of $T_{U\delta}$ = 0.734 and 0.754 at 
\textit{L} = 150 and 80 respectively; (b) An exponential fit 
(dashed curve) to the low temperature data superimposed on the
temperature variation of $d_{z}$ (solid line) shows  proliferation 
of defects at $T_{U\delta} \sim 0.75$ at \textit{L}=80.} 
\label{fig:4}
\end{figure}   
  
  The temperature variation profiles of 
 ($d_{x}, d_{y}, d_{z}$) at different sizes (shown in Fig.~\ref{fig:4}),  
 essentially collapse to a single curve,
except in a very small region near the transition temperature 
$T_{U\delta}$ $(\simeq 0.727)$.  
 In a topological medium these defects are thermally excited 
 at low temperatures below the unbinding transition (leading to their
  exponential growth) and proliferate at its onset
 \cite{Kenna}. Inset(a) of Fig.~\ref{fig:4} magnifies 
 temperature variation of the density to depict the slight size-dependence
 observed near the transition region (shown at $L = 80, 150$).  
 The defect density variation at \textit{L}=80 in this temperature 
 region, along with an exponential fit to data 
 limited to low temperatures ($ \leqslant T_{U\delta}(L = 80)$=0.754), 
 is shown in inset (b), evidencing the onset of proliferation starting 
 near the corresponding unbinding transition temperature. 
 
\begin{figure}
\centering
\includegraphics[width=0.45\textwidth]{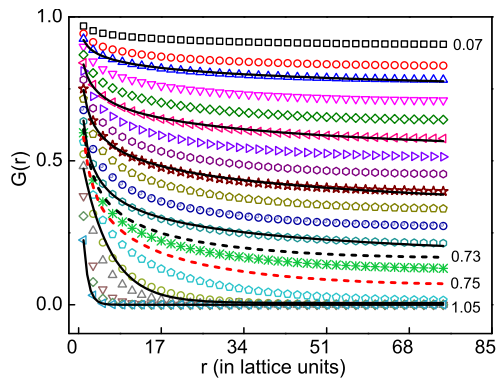}
\caption{(color online) Spatial variation of correlation functions 
$G(r)$ at representative temperatures bracketing the transition
(\textit{L} = 150). $G(r)$ fits well to exponential 
decays above $T$ = 0.75, while it exhibits a power law variation
below $T$ = 0.73. These bounding decays are indicated in the 
figure as dashed lines. The fit curves both above and below the 
transition temperature are superimposed on the corresponding data 
points as solid lines at a few representative temperatures 
 ($T$ = 0.17, 0.35, 0.53, 0.7, 0.78 and 1.05).}
\label{fig:5}
\end{figure}     
 
Spatial variations of $G(r)$ at \textit{L}=150, (depicting only a subset
 of the data computed at 80 temperatures), are shown in Fig.~\ref{fig:5}.
Each profile represents identical variation of the three 
directors. For $T \leq 0.73$, the correlation functions obey power law 
decays very well $G(r,T)\approx r^{-\eta(T)}$, yielding a
temperature dependent exponent $\eta(T)$ (within 1$ \%$ error).
 $\eta(T)$ is found to vanish as $T \rightarrow 0$,
 fitting very satisfactorily to the expression   
$\eta(T)= B (T_{U\eta} - T)^{\kappa}+ \eta_{T_{U}}$, yielding 
$T_{U\eta}(L=150) = 0.729 \pm 0.001$, $\kappa = 0.485 \pm 0.005$, 
$\eta_{T_{U}} = 0.342 \pm 0.003$, and $B = 0.399 \pm 0.002$.
Here  $T_{U\eta}$ is an estimate of the unbinding transition 
temperature derived from the  correlation function data in the low
temperature region, comparing well within errors with the corresponding 
$T_{U\delta}$ (=0.734) at \textit{L} = 150. $\eta_{T_{U}}$ is the
 asymptotic value of the exponent at the transition, and $\kappa$ 
 is the exponent  quantifying vanishing of  $\eta(T)$ as 
 $T\rightarrow 0$. The $G(r)$ profiles for $T\geq 0.75$  fit 
 very well to decays given by, 
$G(r)=A\ r^{-\eta_{T_{U}}} \ \exp \left[-r/\xi \right]+A_{0}$, assigning system 
length scales $\xi(T)$ (within $2 \%$ error) originating from 
correlations limited by the unbounded defect density in the disordered
 state. Here A is a non-universal constant, $A_{0}$ is related to 
 long-range orientational order, and $\eta_{T_{U}}$ is known from the 
 low temperature data. In the small temperature range [0.73 to 0.75], 
 functional dependence of $G(r)$ could not be assigned
satisfactorily to either of the above decay functions 
with relatively much higher least-square errors, and this region is
indicated by two decays in dashes in  Fig.~\ref{fig:5}. 
 
\begin{figure}
\centering
\includegraphics[width=0.45\textwidth]{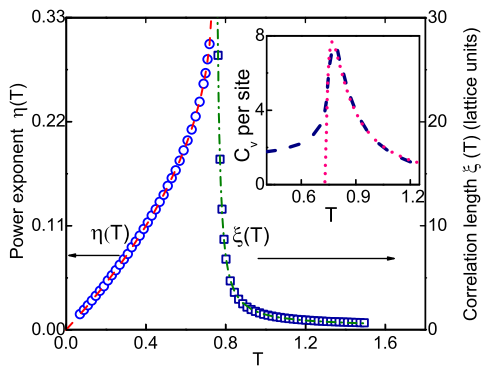}
\caption{(color online) Variation of the exponent $\eta$(T) and
$\xi$(T) with temperature at  \textit{L} = 150. The dashed
line (left) is the power law fit to $\eta$(T) and the dash-dotted 
line (right) is a fit to $\xi$(T) as indicated in the text. Inset 
shows the critical contribution to the $C_{v}$ (short dashes)
superposed  on  the $C_{v}$ profile (long dashes). The contribution
drops to zero  at the temperature $T \approx$ 0.727 expected of an
essential singularity at the transition.} 
\label{fig:6}
\end{figure}   

Fig.~\ref{fig:6} plots the variations 
of $\eta(T)$ and $\xi(T)$ on two abscissae with common temperature 
ordinate. With the estimated $T_{U\eta}$ value, the divergence of 
the  correlation lengths in the disordered state are fit to 
$ \xi(T)\approx \exp \ \left[\frac{D}{(T-T_{U\eta})^{\nu}}\right]$
\cite{Kosterlitz}, obtaining $\nu = 0.304 \pm 0.004$.
With these best fit values (at $L$ = 150) of $T_{U\eta}$ and 
$\nu$ obtained from the $G(r,T)$ data, critical contribution 
(arising from the unbinding of defects) to the specific heat 
profile (together with a background component), given by
$C_{v} \approx \left(\frac{C}{(T - T_{U\eta})}\right)^{2(\nu +1)} 
\ \exp \left[-2 \left(\frac{C}{(T - T_{U\eta})}\right)^{\nu} \right]$
 \cite{Kawamura,Kenna} is fit to the variation obtained from MC simulation 
 at \textit{L} = 150 (Fig.~\ref{fig:1}). The fit parameters are 
 non-universal constants. We note that the critical contribution 
 vanishes as an essential weak singularity at the unbinding temperature. 
 The MC simulated $C_{v}$ (per site) profile (Fig.~\ref{fig:1}) is 
 repeated as inset in Fig.~\ref{fig:6} (long dashes),  plotted along 
 with critical contribution calculated with the above expression, 
 depicted as short dashes. The critical value of $C_{v}$ is seen to 
 expectedly drop to zero abruptly  near the transition temperature. 
 Critical parameters  ($T_{U\eta}, \nu$; from Fig.~\ref{fig:6}) derived 
 from $G(r,T)$  could fairly convincingly generate the observed 
 energy fluctuations above the critical point, based on a 
 topological model of defect-mediated mechanism for the transition.
The power law of $G(r,T)$  below 
$T_{U\eta} (\approx T_{U\delta} )$ = 0.729 shows a quasi long
range order of the medium, - a critical state.

The present values of $\eta_{T_{U}}$ (= 0.342) and $\nu$ (= 0.304) 
 differ from the mean-field values of the 2d \textit{XY}
 model, $\eta_{c}$ (= 0.25) and $\nu$ (= 0.5). This is to be  
 expected owing to fundamental differences in their $\mathcal{R}$ 
 space topology and  $\Pi_{1}(\mathcal{R})$ groups. Such 
 numerical deviations were observed earlier in systems with 
 $\mathcal{R} \cong RP^{2}$ ($Z_{2}$ site symmetry 
 with \textit{n} = 3) like the case with numerical studies on 
 uniaxial 2d LC model reporting $\eta_{c}$ = 0.338 \cite{Botet}. 
 Similar argument was advanced to account for this discrepancy 
 in two-dimensional fully frustrated anti-ferromagnetic Heisenberg 
 model on triangular lattice \cite{Kawamura}. 
 
The results point to the circumstance that, while apolar order is a 
prerequisite to sustain stable topological defects in \textit{n}=3 
model, an enriched order parameter topology engaging all the spin degrees 
in the formation of distinct defects, backed by suitably chosen 
Hamiltonian model, is necessary to facilitate their participation in 
the successful mediation of a topological transition to a critical state. 
Such choices of topology, requiring a non-Abelian fundamental group, 
avoid possible onset of other intervening mechanisms, as was encountered 
in the case of the model with a single class of defects \cite{BKLPRL}. 
        
 We acknowledge the computational support from the Centre for  Modelling
 Simulation and Design (CMSD) and the School of Computer and 
Information Sciences (DST PURSE - II Grant) at the University of
 Hyderabad. BKL acknowledges financial support from Department of
 Science and Technology, Government of India  vide grant ref No: 
 SR/WOS-A/PM-2/2016 (WSS) to carry out this work.

\end{document}